\documentclass[twoside,twocolumn,9pt]{article}
\usepackage{extsizes}
\usepackage[super,sort&compress,comma]{natbib} 
\usepackage[version=3]{mhchem}
\usepackage[left=1.5cm, right=1.5cm, top=1.785cm, bottom=2.0cm]{geometry}
\usepackage{balance}
\usepackage{mathptmx}
\usepackage{sectsty}
\usepackage{graphicx} 
\usepackage{lastpage}
\usepackage[format=plain,justification=justified,singlelinecheck=false,font={stretch=1.125,small,sf},labelfont=bf,labelsep=space]{caption}
\usepackage{float}
\usepackage{fancyhdr}
\usepackage{fnpos}
\usepackage[english]{babel}
\addto{\captionsenglish}{%
  
}
\usepackage{array}
\usepackage{droidsans}
\usepackage{charter}
\usepackage[T1]{fontenc}
\usepackage[usenames,dvipsnames]{xcolor}
\usepackage{setspace}
\usepackage[compact]{titlesec}
\usepackage{hyperref}

\usepackage{import}
\usepackage{amsmath}
\usepackage{amssymb}
\usepackage{bbold}
\usepackage{graphicx}
\usepackage{multicol}
\usepackage{multirow}
\usepackage{adjustbox}
\usepackage{comment}

\definecolor{cream}{RGB}{222,217,201}

\begin{document}

\pagestyle{fancy}
\thispagestyle{plain}
\fancypagestyle{plain}{
\renewcommand{\headrulewidth}{0pt}
}

\newcommand{\revision}[1]{\textcolor{black}{{#1}}}
\newcommand{\newrevision}[1]{\textcolor{black}{{#1}}}


\makeFNbottom
\makeatletter
\renewcommand\LARGE{\@setfontsize\LARGE{15pt}{17}}
\renewcommand\Large{\@setfontsize\Large{12pt}{14}}
\renewcommand\large{\@setfontsize\large{10pt}{12}}
\renewcommand\footnotesize{\@setfontsize\footnotesize{7pt}{10}}
\makeatother

\renewcommand{\thefootnote}{\fnsymbol{footnote}}
\renewcommand\footnoterule{\vspace*{1pt}%
\color{cream}\hrule width 3.5in height 0.4pt \color{black}\vspace*{5pt}} 
\setcounter{secnumdepth}{5}

\makeatletter 
\renewcommand\@biblabel[1]{#1}            
\renewcommand\@makefntext[1]%
{\noindent\makebox[0pt][r]{\@thefnmark\,}#1}
\makeatother 
\renewcommand{\figurename}{\small{Fig.}~}
\sectionfont{\sffamily\Large}
\subsectionfont{\normalsize}
\subsubsectionfont{\bf}
\setstretch{1.125} 
\setlength{\skip\footins}{0.8cm}
\setlength{\footnotesep}{0.25cm}
\setlength{\jot}{10pt}
\titlespacing*{\section}{0pt}{4pt}{4pt}
\titlespacing*{\subsection}{0pt}{15pt}{1pt}

\fancyfoot{}
\fancyfoot[LO,RE]{\vspace{-7.1pt}\includegraphics[height=9pt]{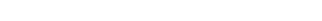}}
\fancyfoot[CO]{\vspace{-7.1pt}\hspace{13.2cm}\includegraphics{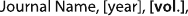}}
\fancyfoot[CE]{\vspace{-7.2pt}\hspace{-14.2cm}\includegraphics{RF}}
\fancyfoot[RO]{\footnotesize{\sffamily{1--\pageref{LastPage} ~\textbar  \hspace{2pt}\thepage}}}
\fancyfoot[LE]{\footnotesize{\sffamily{\thepage~\textbar\hspace{3.45cm} 1--\pageref{LastPage}}}}
\fancyhead{}
\renewcommand{\headrulewidth}{0pt} 
\renewcommand{\footrulewidth}{0pt}
\setlength{\arrayrulewidth}{1pt}
\setlength{\columnsep}{6.5mm}
\setlength\bibsep{1pt}

\makeatletter 
\newlength{\figrulesep} 
\setlength{\figrulesep}{0.5\textfloatsep} 

\newcommand{\topfigrule}{\vspace*{-1pt}%
\noindent{\color{cream}\rule[-\figrulesep]{\columnwidth}{1.5pt}} }

\newcommand{\botfigrule}{\vspace*{-2pt}%
\noindent{\color{cream}\rule[\figrulesep]{\columnwidth}{1.5pt}} }

\newcommand{\dblfigrule}{\vspace*{-1pt}%
\noindent{\color{cream}\rule[-\figrulesep]{\textwidth}{1.5pt}} }

\makeatother

\twocolumn[
  \begin{@twocolumnfalse}
{\includegraphics[height=30pt]{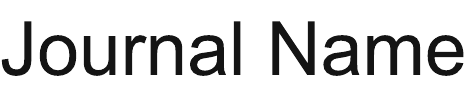}\hfill\raisebox{0pt}[0pt][0pt]{\includegraphics[height=55pt]{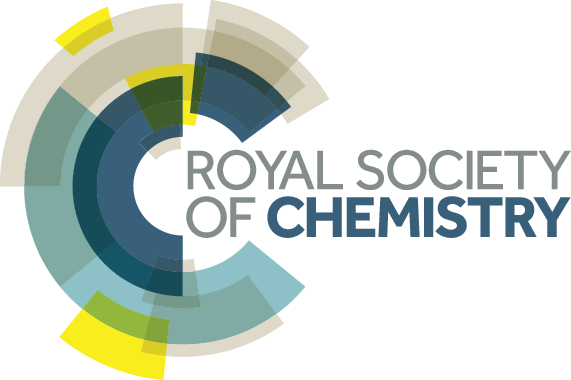}}\\[1ex]
\includegraphics[width=18.5cm]{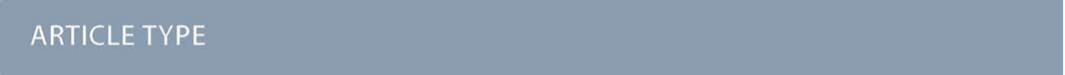}}\par
\vspace{1em}
\sffamily
\begin{tabular}{m{4.5cm} p{13.5cm} }

\includegraphics{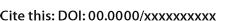} & \noindent\LARGE{\textbf{Cell sorting by active forces in a phase-field model of cell monolayers$^\dag$}} \\
\vspace{0.3cm} & \vspace{0.3cm} \\

 & \noindent\large{James N. Graham,\textit{$^{a,\ddag}$} Guanming Zhang,\textit{$^{b,}$}\textit{$^{c}$} and Julia M. Yeomans\textit{$^{a}$}} \\

\includegraphics{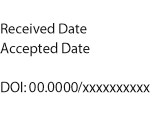} & \noindent\normalsize{Cell sorting, the segregation of cells with different properties into distinct domains, is a key phenomenon in biological processes such as embryogenesis. We use a phase-field model of a confluent cell layer to study the role of activity in cell sorting. We find that a mixture of cells with extensile or contractile dipolar activity, and which are identical apart from their activity, \newrevision{quickly} sort into small, elongated patches \newrevision{which then grow slowly in time}. We interpret the sorting as driven by the different diffusivity of the extensile and contractile cells, mirroring the ordering of Brownian particles connected to different hot and cold thermostats. We check that the free energy is not changed by either partial or complete sorting, thus confirming that activity can be responsible for the ordering even in the absence of thermodynamic mechanisms.} \\

\end{tabular}

 \end{@twocolumnfalse} \vspace{0.6cm}

  ]

\renewcommand*\rmdefault{bch}\normalfont\upshape
\rmfamily
\section*{}
\vspace{-1cm}


\footnotetext{\textit{$^{a}$~Rudolf Peierls Centre for Theoretical Physics, Parks Road, University of Oxford, Oxford, OX1 3PU, UK}}
\footnotetext{\textit{$^{b}$~Center for Soft Matter Research, Department of Physics, New York University, New York 10003, USA}}
\footnotetext{\textit{$^{c}$~Simons Center for Computational Physical Chemistry, Department of Chemistry, New York University, New York 10003, USA}}

\footnotetext{\dag~Electronic Supplementary Information (ESI) available: [details of any supplementary information available should be included here]. See DOI: 00.0000/00000000.}
\footnotetext{\textit{$^{\ddag}$~E-mail: james.graham@physics.ox.ac.uk}}




\section{Introduction}


Cell sorting, the separation of mixtures of different cell-types into distinct domains reminiscent of phase ordering in fluids, has long been a topic of interest in the biological literature \cite{moscona:1952, steinberg:1962,krens:2011}. It is a vital component of embryogenesis and morphogenesis, relevant to how cells organise into different cell types prior to differentiation, and is conserved across vertebrates and invertebrates. Model systems, for example plated confluent layers which include different cell types also often show localised de-mixing \cite{steinberg:1962,balasubramaniam:2021,suzuki:2021}. A number of mechanisms have been suggested to explain cell sorting in tissues, and investigated both \textit{in vitro} and \textit{in silico}. These can appeal to differences in cell adhesion, line tension, or activity \cite{foty:2005,landsberg:2009,skamrahl:2022}. Other characteristics of a monolayer such as cell motility \cite{heine:2021} or area and shape \cite{sahu:2020} may also play a role in cell sorting.

The Differential Adhesion Hypothesis introduced by Steinberg \cite{steinberg:1962} proposes that a cell preferentially adheres to cells of the same, rather than another, type. Differential adhesion leads to variation in tissue surface tension, and results in a thermodynamic separation akin to that of oil and water. The earliest mechanism proposed to explain cell sorting, the DAH gained currency and has been investigated experimentally in cell aggregates expressing varying levels of cadherins \cite{foty:2005} . The hypothesis has been used to describe the sorting of heterogeneous monolayers \textit{in vitro} \cite{skamrahl:2022}, and \textit{in silico} through agent-based and cellular-Potts model simulations \cite{taylor:2012,durand:2021}.

Complementary to the DAH is the Differential Interfacial Tension Hypothesis, which considers the tension at cell-cell junctions. This hypothesis holds that line tension can modulate and regulate cell sorting, and even permit one tissue to completely surround another \cite{brodland:2002}. It has been shown that cell junction tension controls cell sorting in \textit{Drosophila} \cite{landsberg:2009}. In addition, fluctuations in line tension in a vertex model can result in intercalations \cite{sknepnek:2022} and cell sorting \cite{krajnc:2020}.

Experiments by Skamrahl et al. show that differences in the line tension at junctions between cells of different types, which the authors refer to as contractility, has an interplay with differential adhesion during cell sorting \cite{skamrahl:2022}. In this work, the authors compare cultures of wild-type Madin-Darby Canine Kidney cells to `dKD' cells, which have a protein knockdown that enhances contractility. To analyse cell sorting, they tag wild-type cells with green fluorescent protein and mix them with either dKD cells or untagged wild-type cells. The mixture of dKD and wild-type cells separates into patches of each type. The authors also observe that demixing proceeds in a two-stage process: a fast stage followed by a slow stage. They attribute the fast time scale to differential contractility, and the slow time scale to differential adhesion.

Other factors have also been implicated in cell sorting in a monolayer. Heine et al. observe the influence of the solid- or fluid-like nature of a cell culture on sorting within the differential adhesion framework \cite{heine:2021}. There are also suggestions that the apical surface area of epithelial \revision{cells} can affect sorting \cite{skamrahl:2022,sahu:2020}.

Moreover, living systems are \textit{active}. They continuously use chemical energy to drive motility and growth, and hence remain out of thermodynamic equilibrium. Recent work has successfully interpreted collective processes in mechanobiology in terms of the theories of active matter, and in particular of active nematics. Balasubramaniam et al. \cite{balasubramaniam:2021} observed incomplete cell sorting in a mixture of two MDCK cell strains with different strengths of inter-cellular interactions. They interpreted these results in terms of a mixture of dipolar active extensile and active contractile cells and complemented the experimental results with simulations of a phase-field model. A two-fluid continuum model has also been used to show that fluids of different activities can undergo micro-phase separation \cite{bhattacharyya:2023}.

Thus many different mechanisms can result in a degree of cell sorting, and  the interplay of several may be necessary for complete, well-controlled segregation in a given biological process \cite{suzuki:2021,heine:2021}. To obtain a more complete understanding of sorting phenomena in cell monolayers and tissues, it is crucial to investigate each of the mechanisms involved. Isolating individual processes is difficult in experiment but modelling can play a useful role in this regard. Therefore here we implement a phase-field model and concentrate on asking whether active forces acting between cells, which arise from cadherin-based intercellular junctions, can result in sorting in a confluent cell monolayer.

In the next section, we introduce the phase-field model, including both the passive dynamics of each cell and the active forces between cells. In section 3 we present results indicating that cells of different activities undergo localised sorting. Our results are summarised and discussed in section 4.

\section{Phase-Field Model}

Phase-field models are simplified approaches to cellular dynamics \cite{nonomura:2012, najem:2016, mueller:2019, mouregomez:2019, camley:2017, alert:2020} which have been used to model the motility of single  and multiple cells and of confluent cell layers.  A particular advantage of phase-field approaches is that it is possible to distinguish passive and active contributions to cell dynamics, and to implement both intracellular forces and forces between cells mediated through cell junctions in a transparent way.

Each cell $i$ is represented by a phase field, $\varphi^{\revision{(i)}}(\mathbf{x})$. The phase field takes a value 0 outside the cell and 1 in the interior and varies smoothly across the cell boundary. The phase field $\varphi^{\revision{(i)}}(\mathbf{x})$ moves with a velocity $\mathbf{v}^{\revision{(i)}}(\mathbf{x})$, and evolves according to
\begin{equation}
\label{eq:dynamics}
\partial_t \varphi^{(i)}(\mathbf{x}) + \mathbf{v}^{(i)} (\mathbf{x}) \cdot \nabla \varphi^{(i)}(\mathbf{x} ) = -J_0 \frac{\delta {\mathcal{F}}}{\delta \varphi^{(i)} (\mathbf{x})}
\end{equation}
where $\mathcal{F}$ is a free energy. The right-hand side of \revision{E}quation~(\ref{eq:dynamics})
describes the dissipative relaxation dynamics of the cells to a free energy minimum at a rate $J_0$. 
Assuming over-damped dynamics, the velocity $\mathbf{v}^{(i)} (\mathbf{x})$ of a cell is determined by the local force density acting on the cell,
\begin{equation}
\label{eq:force_balance}
\xi \mathbf{v}^{(i)} (\mathbf{x}) =\mathbf{f}_{\text{passive}}^{(i)} (\mathbf{x}) + \mathbf{f}_{\text{active}}^{(i)} (\mathbf{x})
\end{equation}
where we distinguish active and passive contributions to the force density and $\xi$ is a friction coefficient.
\revision{The force balance holds everywhere in the domain: any force imbalance is countered by the dynamic friction which arises from substrate interactions, \newrevision{representing} focal adhesions \cite{ladoux:2017}.} The passive force density
\begin{equation}
\mathbf{f}^{(i)}_{passive}(\mathbf{x})=\frac{\delta\mathcal{F}}{\delta\varphi^{(i)}}\nabla\varphi^{(i)}\qquad
\end{equation}
is modelled by three free energy contributions:
\begin{subequations}\label{eqn:f}
\begin{align}
\mathcal{F}_{CH}&=\sum_i\frac{\gamma}{\lambda}\int\textrm{d}\mathbf{x}\left[\varphi^{(i)}(\mathbf{x})^2(1-\varphi^{(i)}(\mathbf{x}))^2+\lambda^2(\nabla\varphi^{(i)}(\mathbf{x}))^2\right]\label{ch},\\
\mathcal{F}_{area}&=\sum_i\mu\left[1-\frac{1}{\pi R^2}\int\textrm{d}\mathbf{x}\,\varphi^{(i)}(\mathbf{x})^2\right]^2\label{area},\\
\mathcal{F}_{rep}&=\sum_i\sum_{j\neq i}\frac{\kappa}{\lambda}\int\textrm{d}\mathbf{x}\,\varphi^{(i)}(\mathbf{x})^2\varphi^{(j)}(\mathbf{x})^2.
\end{align}
\end{subequations}
The first of these is a Cahn-Hilliard term, which favours phase separation into regions of $\varphi^{(i)}=0,1$ with an interface \revision{width} $\lambda$. \revision{The parameter $\gamma$ controls the line tension for the interface of a single cell, which \newrevision{governs} how easily the cell can deform.} The second free energy term imposes a soft area constraint with energy scale $\mu$, where $\pi R^2$ is the equilibrium area of a cell and $R$ its target radius. The third term penalises cell-cell overlap with an energy scale $\kappa$. The length unit in the system is equal to the lattice unit

It is next necessary to define the active forces acting in the system. The polar motility of individual epithelial cells arises from lamellipodia and other protrusions \cite{rorth:2009, mayor:2016} that generate polar forces. However, it has been argued that the formation of lamellipodia is suppressed in confluent epithelia, a phenomenon known as contact inhibition of locomotion \cite{carmona-fontaine:2008}, suggesting a reduction in the strength of any persistent polar force in the monolayer. There is evidence that individual epithelial cells can create dipolar stresses \cite{balasubramaniam:2021}, and many cell monolayers exhibit active flows very similar to those predicted by active nematic models \cite{mueller:2019}. Therefore we choose to concentrate on dipolar (balanced) active forces.

To calculate the active dipolar contribution to the force density, $\mathbf{f}_{\text{active}}^{(i)} (\mathbf{x})$, we first calculate the deformation tensor that quantifies the shape of each cell \cite{mueller:2019, bigun:1987}, 
\begin{align}
\label{eq:deformation_tensor}
\begin{split}
{\mathcal D}^{(i)}
&= - \int d\mathbf{x} \left\{\nabla \varphi^{(i)} \nabla \varphi^{(i)^T} - {\frac{\mathbb{1}}{2}} \mbox{Tr}( \nabla \varphi^{(i)} \nabla \varphi^{(i)^T} )\right\} \\
& \equiv \sqrt{({\mathcal D}^{(i)}_{xx})^2 + ({\mathcal D}^{(i)}_{xy})^2}\; (\mathbf{d}_{\parallel}^{(i)} \mathbf{d}_{\parallel}^{(i)^T} -\mathbf{d}_{\bot}^{(i)} \mathbf{d}_{\bot}^{(i)^T})
\end{split}
\end{align}
where $\mathbf{d}_{\parallel}^{(i)}$ and $\mathbf{d}_{\bot}^{(i)}$ are the orthonormal eigenvectors of the ${\mathcal D}^{(i)}$, along and perpendicular to the elongation axis of the cell respectively, normalised so that 
$\mathbf{d}_{\parallel}^{(i)}\mathbf{d}_{\parallel}^{(i)^T} + \mathbf{d}_{\bot}^{(i)}\mathbf{d}_{\bot}^{(i)^T} = \mathbb{1}$.

We next define a director, $ \mathbf{n}^{(i)}$, associated with each cell $i$ and assume that $\mathbf{n}^{(i)}$ relaxes towards  $\mathbf{d}_{\parallel}^{(i)}$ 
\begin{equation}
\frac{\mathrm{d}\mathbf{n}^{(i)}}{\mathrm{d}t} = -J_n(\mathbf{n}^{(i)}-\mathbf{d}_{\parallel}^{(i)}) 
\label{eq:fluctuation-dynamics}
\end{equation}
where $J_n$ controls the time scale of relaxation. We assume no noise in the relaxation of the director. 

In the absence of any unbalanced active forces, the dipolar contribution to the active stress acting on cell $i$ is split into a self-induced stress and a stress due to the cell's neighbours, related to the director field by \cite{simha:2002,zhang:2021}
\begin{equation}
\sigma_{\alpha \beta}^{(i)}(\mathbf{x})=
-\zeta^{(i)}_{\text{self}}Q^{(i)}_{\alpha\beta}\varphi^{(i)}(\mathbf{x})- \sum_{ j \neq i} \zeta^{(j)}_{\text{inter}}Q_{\alpha \beta}^{(j)} \;\varphi^{(j)}(\mathbf{x}) 
\label{eq:inter-stress}
\end{equation}
where each cell is assigned a nematic tensor
\begin{equation}
Q_{\alpha \beta}^{(i)} 
= \revision{n}^{(i)}_\alpha \revision{n}^{(i)}_\beta - \frac{|\revision{\mathbf{n}}^{(i)}|^2}{2}  \delta_{\alpha \beta}.
\label{eq:Q-tensor}
\end{equation}
The force density arising from the active stress is then
\begin{equation}
\mathbf{f}_{\text{active}}^{(i)}(\mathbf{x}) = \nabla \cdot \mathbf{\sigma}^{(i)} 
   =-\zeta^{(i)}_{\text{self}}\mathbf{Q^{(i)}}\cdot\nabla\varphi^{(i)}(\mathbf{x})-\sum_{ j \neq i}  \zeta^{(j)}_\text{{inter}}\mathbf{Q}^{(j)} \cdot \nabla \varphi^{(j)}(\mathbf{x}).
   \label{eq:active-force}
\end{equation}
Biologically, $\zeta_{\text{self}}$ is a measure of the magnitude of stresses linked to intracellular myosin motors, while $\zeta_{\text{inter}}$ indicates the strength of intercellular stresses mediated by adherens junctions acting between neighbouring cell cortices.

The parameters assigned to the free energy terms (\ref{eqn:f}) were $\gamma=1.4$, $\lambda=2.0$, $\mu=120$ and $\kappa=1.5$. The coefficient of friction in \revision{E}quation~(\ref{eq:force_balance}) was $\xi=3.0$ and the target cell radius was $R=8.0$. Relaxation according to \revision{E}quation~(\ref{eq:dynamics}) was controlled by the rate $J_0=5\times10^{-3}$. Each cell's nematic director relaxed towards its long axis $\mathbf{d}_{\parallel}^{(i)}$ at rate $J_{n}=0.1$. \newrevision{The timescale for the shape of a cell to relax according to the Cahn-Hilliard phase separation term is $(\gamma \Delta\ell J_0)^{-1}\sim10^2$, where $\Delta\ell=1$ is the lattice unit.}

The phase-field model was implemented with periodic boundary conditions. The system was initiated by placing cells randomly within relevant areas of the simulation domain \revision{to give a confluent layer}. The system's initial configuration was obtained by relaxing the randomly-placed cells as though they were passive for $5\times10^3$ time steps, at which point activity was turned on \newrevision{(denoted by $t=0$)} and data \newrevision{were} recorded to $t=\newrevision{1\times}10^5$ \newrevision{or $t=5\times10^5$} timesteps. \revision{The dynamical equations were solved using a forward Euler scheme} \newrevision{with one predictor-corrector step}. \newrevision{The unit of time is 1, with each step sub-divided into five substeps, and the unit of length is the lattice spacing.} \newrevision{We note that the model differs from the first implementation of this software\cite{mueller:2019} by the omission of an adhesive contribution to the free energy and by the coupling of $\mathbf{Q}^{(i)}$ not directly to $\mathcal{D}^{(i)}$ but to its principal eigenvector $\mathbf{d}^{(i)}_\parallel$. Intercellular interactions, treated separately from intracellular dipolar stresses, have been investigated more recently\cite{zhang:2021}, and the treatment here follows this recent work.}

\section{Results}

We \newrevision{concentrate primarily} on dipolar intercellular interactions, choosing $\zeta_{\text{self}}=0$ and returning to consider intracellular forces at the end of this section. To investigate whether cell sorting can be driven by differences in active inter-cellular forces, we prepared an equal mixture of cells with extensile and contractile activities, $\zeta_{\text{inter}}=0.4$ and $\zeta_{\text{inter}}=-0.4$, respectively. All other simulation parameters were identical for both cell types. \newrevision{The simulation was started by placing 672 cells of each type randomly in the simulation domain which measured 560 by 560 lattice units. This number density corresponds} \revision{to a cell density where the cell layer is confluent, but cell motion is still possible.} \newrevision{Although a `packing fraction' is not uniquely defined for soft particles, we describe the target packing fraction as the total area fraction of cells divided by the area of the domain, which is $\sim0.86$.}

\begin{figure}[h!]
\includegraphics[scale=0.55]{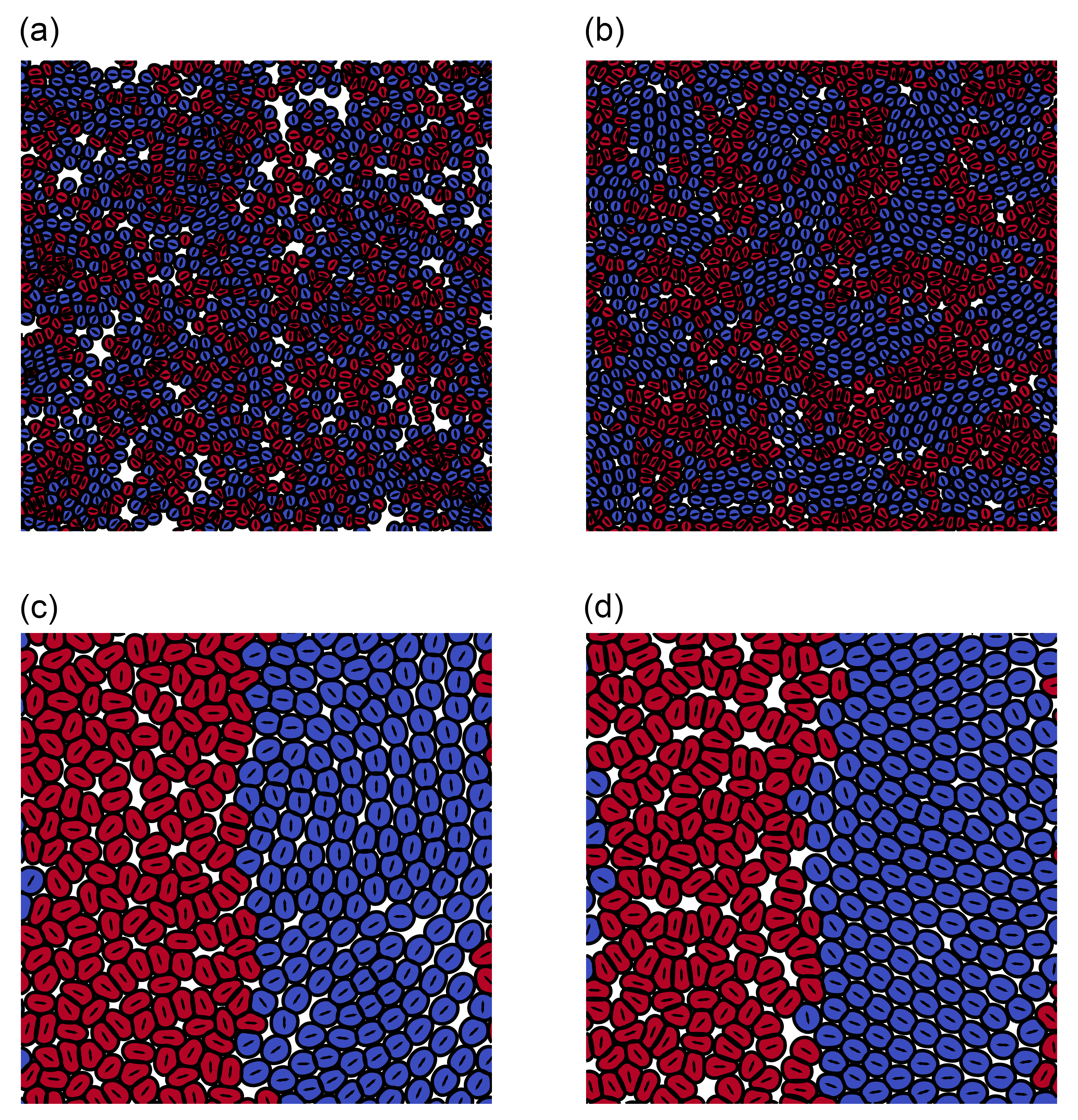}
\caption{Snapshots of segregation in a 1:1 mixture of extensile (red) and contractile (blue) cells. (a) \newrevision{$t=1000$}, (b) \newrevision{$t=5\times10^5$} for a fully mixed initial condition showing  microphase separation. (c) \newrevision{$t=1000$}, (d) $t=\newrevision{1\times}10^5$ for a state that is fully sorted at the beginning of the simulation and which does not mix.   Note that the contractile cells develop nematic order.}
\label{fig:segregation}
\end{figure}

Figure~\ref{fig:segregation}(a) and (b) illustrate the configurations at \newrevision{$t=1000$} and after \newrevision{$t=5\times10^5$} time steps. The snapshots show evidence of partial cell sorting, into elongated clusters\revision{; these clusters coalesce, break up, and re-form during the simulation}. As a comparison we \newrevision{simulate} a fully sorted state \newrevision{with 336 cells in a $280\times280$ box}, with extensile and contractile cells separated into macroscopic regions as shown in Fig.~\ref{fig:segregation} (c), (d). The cells show no tendency to mix on the time scale of the simulation \newrevision{(here to $t=1\times10^5$)}\revision{, therefore both microphase and macrophase separation persist for long times.} Note that contractile cells are elongated slightly by the dipolar forces and demonstrate nematic alignment. \revision{We seek to quantify the phase separation and describe its mechanism.}

There are a number of ways to measure the segregation in a system \cite{balasubramaniam:2021,skamrahl:2022,taylor:2012,mehes:2011}. Following \cite{skamrahl:2022}, we define a segregation index by counting the neighbours of each cell. Two phase-field cells are identified as neighbours when the corresponding phase fields take a value greater than a certain threshold, set here to be 0.1, on the same lattice site. The segregation index is then defined by
\begin{equation}
SI=\left\langle\frac{n}{n+\bar{n}}\right\rangle,
\end{equation}
where $n$ is the number of neighbours of a cell with the same activity, $\bar{n}$ is the number of neighbours of a cell with opposite activity and the average is taken over all cells. With our choice of parameters, the phase-field cells are stable and have sharp interfaces so no spurious neighbour pairs are identified. 

\begin{figure}
\begin{center}
\includegraphics[scale=0.6]{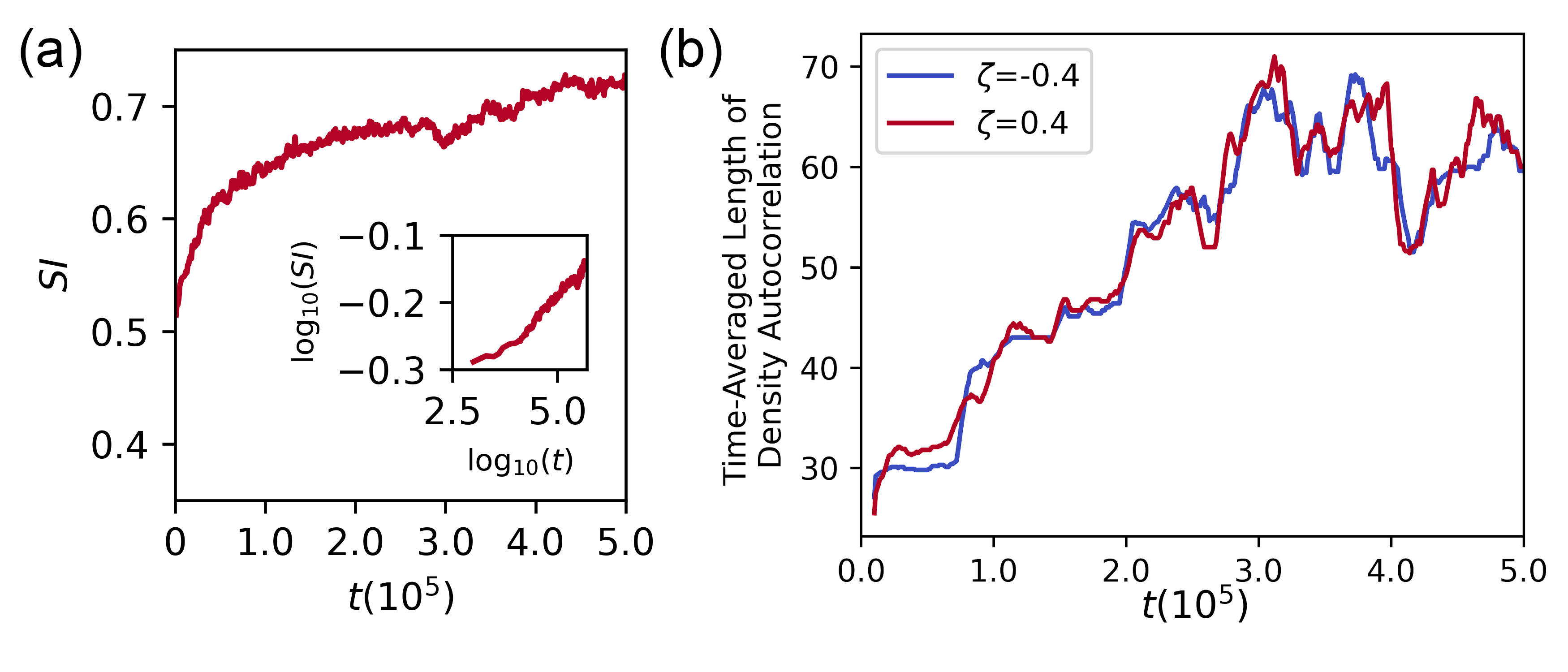}
\end{center}
\caption{Segregation in a 1:1 mixture of extensile and contractile cells. (a) \newrevision{Segregation index $SI$ versus time. The onset of segregation is on a} timescale $t\sim 10^3$\newrevision{, and }. The $SI$ increases \newrevision{past} $\sim 0.7$ on the timescale of the simulation\newrevision{; the inset showing $SI$ on log-log axes indicates the system coarsens steadily.} (b) \newrevision{Time-averaged correlation length of $\rho_\zeta$ for extensile and contractile cells in the simulation illustrated in Fig.~\ref{fig:segregation} (a) and (b). The correlation length grows up to $t\approx3\times10^5$, then appears to saturate on a scale commensurate with the diameter of $\sim4$ cells.}}
\label{fig:corr}
\end{figure}

The segregation index for a 1:1 mixture is plotted in Figure \ref{fig:corr} \newrevision{(a)}. \newrevision{The microphase separation proceeds quickly after the onset of dipolar activity at $t=0$ but then slows. It is unclear from the data whether microphase separation is arrested or continues over very long time scales, since the $SI$ appears to saturate at late times.}

An alternative measure of phase separation is the density autocorrelation function
\begin{equation}
C_\zeta(\mathbf{r}) =\langle\rho_\zeta(\mathbf{x})\rho_\zeta(\mathbf{x}+\mathbf{r})\rangle-\langle\rho_\zeta(\mathbf{x})\rangle\langle\rho_\zeta(\mathbf{x}+\mathbf{r})\rangle
\end{equation}
where the density field $\rho_\zeta$ for the subset of cells in the system with $\zeta_{\text{inter}}=\zeta$ is defined by
\begin{equation}
\rho_\zeta(\mathbf{x}) = \sum_{i: \zeta^{(i)}_{\text{inter}}=\zeta} \revision{\varphi}^{\newrevision{(i)}}(\mathbf{x}).
\end{equation}
\newrevision{Figure~\ref{fig:corr} shows the moving average of the lengthscale of the density autocorrelation $\rho_\zeta$, taken over $10^4$ timesteps.} Recalling that each cell has a nominal diameter $\sim$16 lattice units, the length scale of the density autocorrelation is \newrevision{initially} roughly one cell, consistent with a well-mixed monolayer, while the length scale at long times increases to slightly more than \newrevision{4} cells, consistent with microphase separation into clusters roughly \newrevision{4} cells wide. These results demonstrate quantitatively that the mixture of extensile and contractile cells sorts according to activity. \newrevision{The data suggest that the cluster size saturates, but we cannot rule out a continued extremely slow growth of the clusters. Figure~\ref{fig:corr} (a) is reproduced, alongside plots of $C_\zeta(r)$ at $t=1\times10^3$ and $t=1\times10^5$, in Figure~SI1, while Figure~SI2 provides a comparison of $SI$ and $C(r)$ for a fully sorted system.}

The segregation index and density autocorrelation indicate microphase separation but yield no information about the dynamics of the system. We next examine the mean-square displacement of the phase-field cells, comparing  the fully phase separated state (Fig.~\ref{fig:segregation}d) to the microphase-separated state (Fig.~\ref{fig:segregation}b).

\begin{figure}[h!]
\begin{center}
\includegraphics[scale=0.6]{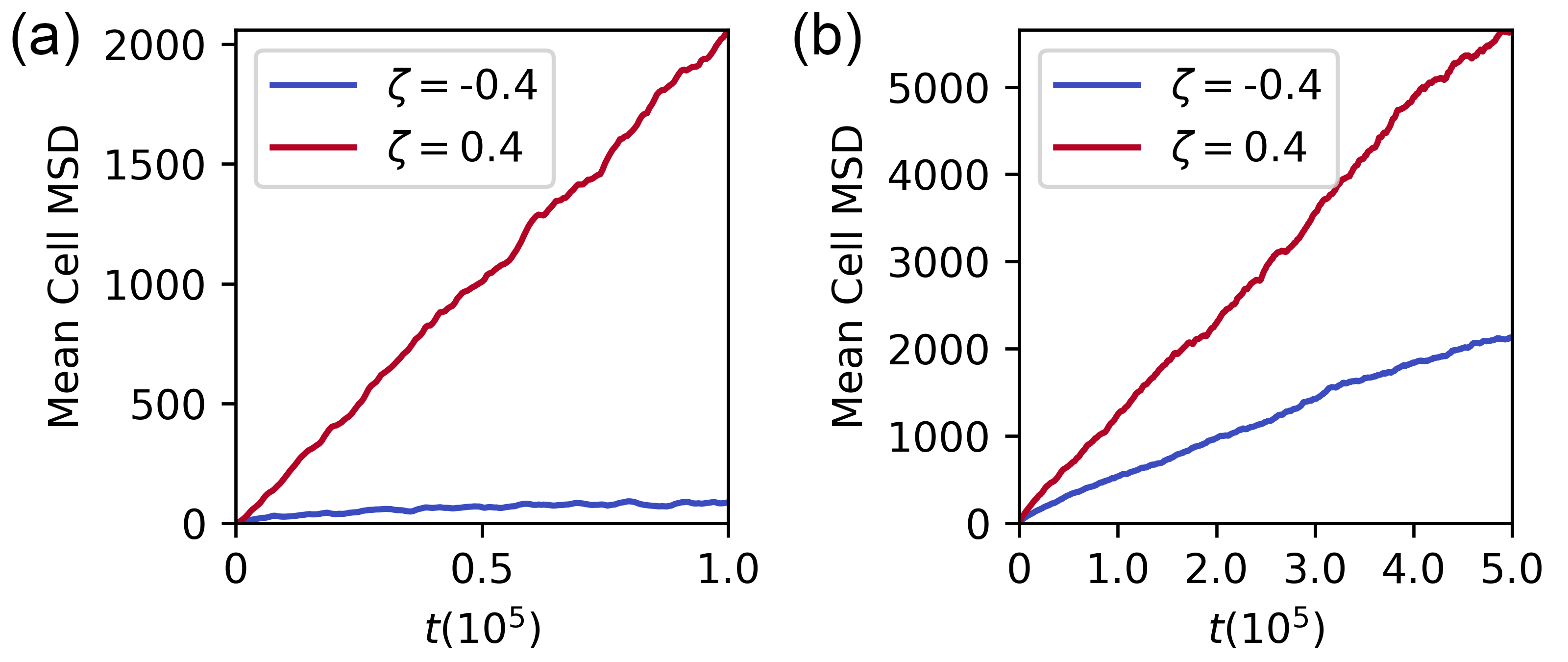}
\end{center}
\caption{Mean-square displacement for contractile (blue) and extensile (red) cells in (a) a 1:1  phase separated monolayer (Fig.~\ref{fig:segregation}d) (b) a 1:1 microphase-separated state (Fig.~\ref{fig:segregation}b). When the cells are completely phase separated, the extensile cells behave as a fluid while the contractile cells behave as a solid; extensile cells diffuse three orders of magnitude as quickly as contractile cells. In the microphase-separated state, however, the presence of contractile cells slows the diffusion of extensile cells, which in turn push the contractile cells around the system. \newrevision{The MSD has not saturated by $t=5\times10^5$, which indicates that the clusters are still evolving and rearranging.}}
\label{fig:msd}
\end{figure}

Results for the phase-separated configuration shown in Fig.~\ref{fig:msd}a indicate that, when surrounded by cells of the same type, the motion of the contractile cells is almost entirely arrested, while extensile cells move more freely. \newrevision{This is true also in pure monolayers: Figures~SI3 and SI4 show, respectively, snapshots of layers and plots of mean-square displacement for homogeneous monolayers of extensile and contractile cells.} This is a result of the different intercellular interactions due to the active dipolar forces. Contractile cells elongate and align to form a solid-like, nematic configuration. Extensile cells prefer to lie at right angles which leads to frustration and gives unstable configurations. Pairs of cells which lie perpendicular to each other have a polarity which results in \revision{net migration} \cite{zhang:2021}.

In the microphase separated system the diffusion of contractile cells is enhanced and that of extensile cells reduced (Fig.~\ref{fig:msd}b). This is because less motile clusters of contractile cells constrain the paths of extensile cells, while the extensile cells tend to push the contractile cells around the simulation domain. \newrevision{As part of the diffusion, individual extensile cells are able to squeeze between contractile clusters from one extensile cluster to another. Given the apparent saturation of the $SI$ and the density autocorrelation lengthscale at long times, the system may be entering a dynamic steady state. It is unclear whether the layer can sort completely, as in Figure~\ref{fig:segregation} (d), at much longer times.}

Taking the diffusion constant in the system to be that of the extensile cells, $\langle\mathbf{x}^2\rangle_{\zeta=+0.4}=4Dt$, the timescale associated with diffusion is $\tau_D=R^2/D\sim 10^4$. \newrevision{Madin-Darby Canine Kidney cells migrate on the order of 10 microns per hour \cite{doxzen:2013}, which is on the order of a cell diameter. The diffusion timescale in the $1:1$ mixed system suggests that $1\times10^5$ timesteps \textit{in silico} corresponds to order a day of real time, which is the duration of the experiments of Balasubramaniam et al. \cite{balasubramaniam:2021}.}

We note that differential diffusivity has been shown to sort otherwise passive soft particles \cite{weber:2016}, even at high packing fractions \cite{mccarthy:2023}. \revision{Here, extensile cells are akin to hot particles and contractile cells to cold ones}. Weber, Weber and Frey \cite{weber:2016} attribute the phase separation in a mixture of particles at two different temperatures to an effective attraction between cold particles; here the active forces act to attract the contractile cells \newrevision{into more coherent clusters}.

\newrevision{In systems that sort, there is often a thermodynamic basis for the phase separation. Consequently, we investigate whether the clustering is associated with any changes in the total free energy of our model, Equations~\ref{eqn:f}. For the microphase separated system (Fig.~\ref{fig:segregation}a,b), the mean and standard deviation of the total free energy per cell from times $t=1\times10^3$ to $t=5\times10^5$ are listed in Table~\ref{table:energy}. As a comparison, data for the system with phase-separated initial condition (Fig.~\ref{fig:segregation}c,d) from times $t=1\times10^3$ to $t=1\times10^5$ are also listed. In both cases, the times run from shortly after the turning on of dipolar active stresses to the end of the simulation. The free energy shows no dependence on time or, indeed, cluster size, giving additional support to an active origin to the cell sorting.}

\begin{table}
\centering
\renewcommand{\arraystretch}{1.2}
\caption{\newrevision{Mean $\bar{\mathcal{F}}_{\textrm{tot}}$ and standard deviation of free energy per cell for the microphase-separated state (Fig.~\ref{fig:segregation}a,b) and the sorted state (Fig.~\ref{fig:segregation}c,d) over time, starting at $t=1\times10^3$, shortly after activity is turned on. The clustering of the phase fields over time is not associated with a reduction in free energy at long times, in contrast to phase separation according to mechanisms such as differential adhesion.}}
\label{table:energy}
\begin{tabular}{*{5}{>{\centering\arraybackslash}p{1cm}}}
\hline
& \multicolumn{2}{p{2cm}}{\centering Microphase-Separated} & \multicolumn{2}{p{2cm}}{\centering Sorted}\\
\hline
$t$ & $\bar{\mathcal{F}}_{\textrm{tot}}$ & $\sigma_{\mathcal{F}}$ & $\bar{\mathcal{F}}_{\textrm{tot}}$ & $\sigma_{\mathcal{F}}$\\
$1\times10^3$ & 39.99 & 4.783 & 38.10 & 2.372\\
$1\times10^4$ & 38.87 & 3.803 & 38.96 & 2.318\\
$5\times10^4$ & 38.88 & 3.824 & 39.19 & 4.028\\
$1\times10^5$ & 38.78 & 3.628 & 38.71 & 3.041\\
$5\times10^5$ & 38.95 & 3.855 & x & x\\
\hline
\end{tabular}
\end{table}
Thus far, we have considered mixtures of extensile and contractile cells, with $\zeta_{\text{inter}}=+0.4, -0.4$. It is possible, in addition, to examine active-passive systems. An extensile-passive system, with $\zeta_{\text{inter}}=+\revision{0.8}, 0.0$, fails to phase separate because the passive cells have no active forces which can hold them together (\newrevision{Figs.~SI5 and SI6}) \newrevision{--- the contractile cells form a solid-like phase in addition to being less diffusive than extensile cells}. A contractile-passive system, with $\zeta_{\text{inter}}=-\revision{0.8}, 0.0$, also fails to phase separate (\newrevision{Figs.~SI7 and SI8}). In this system, contractile cells are not able substantially to deform each other and the geometry remains frustrated. Any randomly-initialised patches of contractile cells adopt a nematic configuration, while any passive cells not subject to contractile stresses remain isotropic. The failure of a mixture of active and passive cells to phase separate is consistent with simulations of active Brownian particles mixed with passive particles \cite{dolai:2018}.

\revision{Finally we mention intracellular dipolar activity $\zeta_{\textrm{self}}\neq0$ \newrevision{(Figs.~SI9, SI10 and SI11)}. As observed in \cite{zhang:2021}, values of $\zeta_{\textrm{self}}$ and $\zeta_{\textrm{inter}}$ with the same sign tend to cancel. This reduction in effective activity reduces the diffusivity of the active phase-field cells and slows sorting.}

\section{Discussion}

There is considerable experimental evidence of sorting of different cell types both \textit{in vitro}, in confluent cell layers \cite{moscona:1952,steinberg:1962,balasubramaniam:2021,skamrahl:2022,mehes:2011} or cellular spheriods \cite{suzuki:2021}, and in vivo, for example during morphogenesis \cite{landsberg:2009,carmona-fontaine:2008}. Many different physical factors are candidates for driving the sorting. Without doubt equilibrium effects, such as the dependence of cell-cell adhesion or line tension of cell-cell junctions on different cell neighbours, can lead to ordering akin to thermodynamically-driven phase ordering in liquid-liquid mixtures \cite{foty:2005,brodland:2002,krajnc:2020}. However, biological systems are naturally out of equilibrium, and it has also been suggested that different forms of activity can result in the sorting of different cell types \cite{balasubramaniam:2021,belmonte:2008}. Balasubramaniam et al. in particular show that cells with a mixture of extensile and contractile force dipoles are able to phase separate on short length scales \cite{balasubramaniam:2021}.

Isolating the different contributions to cell sorting is difficult in experiments, but easier in the context of computational models of cell motility. Therefore in this paper we use a multi-phase field model of a confluent cell layer to study the influence of dipolar active interactions between neighbouring cells on cell sorting. We demonstrate that a mixture of extensile and contractile active dipolar cells, which are otherwise identical, can undergo partial sorting. We interpret this as an out-of-equilibrium effect resulting from the different dynamics of the two cell populations. Extensile cells are smaller and more motile whereas contractile cells tend to elongate and form static\revision{, solid-like} nematic patches. \revision{We caution, however, that both the microphase separated state and a fully sorted state are stable on the timescale of the simulations, and it is unclear whether further coarsening will occur on times we cannot access.}

Any model of cell mechanics must still be viewed with caution as there are still many questions about the model details needed to faithfully reproduce the cells' interactions and dynamics. Here we have focussed on forces, mediated by adherens junctions between cell cortices, that act across cell-cell boundaries, which we have modelled as balanced dipolar forces. However, fluctuating polar forces \cite{vafa:2021} or active forces which act along cell-cell junctions \cite{sknepnek:2022} may be relevant, as may apical-basal asymmetry if a monolayer is modelled in three dimensions \cite{monfared:2023}.

\section*{Conflicts of interest}
There are no conflicts to declare.

\section*{\newrevision{Code availability}}
\newrevision{The software Celadro, used for the simulations in this work, which was written by Romain Mueller, is available on request to the authors.}



\balance


\bibliography{llps-format} 

\end{document}